\documentstyle[11pt,epsfig]{article}

\def\laq{~\raise 0.4ex\hbox{$<$}\kern -0.8em\lower 0.62
ex\hbox{$\sim$}~}
\def\gaq{~\raise 0.4ex\hbox{$>$}\kern -0.7em\lower 0.62
ex\hbox{$\sim$}~}

\def\beq{\begin{equation}}
\def\eeq{\end{equation}}
\def\bea{\begin{eqnarray}}
\def\eea{\end{eqnarray}}
\def\bean{\begin{eqnarray*}}
\def\eean{\end{eqnarray*}}

\def \tb {\bar{t}}
\def \pa {\partial}

\def \ti {\widetilde}

\def \b {\beta}
\def \a {\alpha}

\def \ep {\epsilon}

\def \Om {\Omega}
\def \noi {\noindent}

    \def\be{\begin{equation}}
    \def\ee{\end{equation}}
    \def\ba{\begin{eqnarray}}
    \def\ea{\end{eqnarray}}

    \def\e{\mbox{e}}

    \def\alphat{\tilde{\alpha}}
    \def\betat{\tilde{\beta}}

\begin{document}

\begin{titlepage}

\vspace{0.5cm}

\begin{center}

\huge
 {The cosmological backreaction: gauge (in)dependence, observers and scalars}

\vspace{0.8cm}

\large{G. Marozzi$^{1,2}$}

\normalsize

\vspace{0.5cm}

{\sl $^1$Coll\`ege de France, 11 Place M. Berthelot, 75005 Paris, France}

\vspace{.1in}

{\sl $^2$ GR$\epsilon$CO – Institut d'Astrophysique de Paris, UMR7095 CNRS, \\ 
 Universit\'e Pierre \& Marie Curie, 98 bis boulevard Arago, 75014 Paris, 
France}

\vspace{.1in}

\vspace*{1cm}

\begin{abstract}
\noi
We discuss several issues related to 
a recent proposal for defining classical spatial averages to be used in 
the so-called cosmological backreaction problem.
In the large averaging-volume limit all  gauge dependence disappears 
and different  averages  can be univocally characterized by the observers associated
with  different  scalar fields.
The relation between such averaging procedure and the standard one is emphasized and 
a gauge invariant way to select different observers is presented.
For finite averaging volumes we show that, within our  proposal, a residual gauge 
dependence is left, but  is suppressed by several effects.
\end{abstract}
\end{center}

\end{titlepage}

\newpage

\parskip 0.2cm


\section{Introduction}
\label{Sec1}
\setcounter{equation}{0}

The geometric properties of the Universe, on sufficiently large scale of
distances, may be described by a homogeneous and isotropic 
Friedman-Lemaitre-Robertson-Walker (FLRW) metric.
On the other hand on small scales the Universe exhibits a very inhomogeneous
structure. 
In this way the FLRW metric should be interpreted as 
an ``averaged'' cosmological metric. A problem (see, for
example, \cite{Ellis}) appears due to the fact that the Einstein equations 
for the {\em averaged
geometry} are different, in general, from the {\em averaged Einstein
equations}. In fact the averaging procedure does not commute, in
general, with the non-linear differential operators appearing in the
Einstein equations. As a consequence, the dynamics of the averaged
geometry is affected by so-called ``backreaction'' terms,
originating from the dynamic contribution of the inhomogeneities present
in the metric and matter 
sectors. 

Following the discovery of cosmic acceleration on large
scales, interest in  
these issues has considerably risen (see \cite{Buchert_Review} for a review). 
In fact, it has been suggested (see e.g. \cite{3ref}) 
that the dynamical effects
of the backreaction could replace 
dark-energy sources in the explanation of such cosmic acceleration,
solving, in this way, also the well-known ``coincidence problem''.

In order to have a physical interpretation the results  that one obtains in the investigation of the backreaction
problem should be
independent from the gauge chosen.
More generally, the gauge issue consist in  extracting   
 physically meaningful
results from cosmological perturbation theory eliminating all
possible  gauge artifacts. At linear order in the perturbations, the way to do
this is well known:  following Bardeen's pioneering work \cite{Bardeen}, it
consists of defining gauge invariant combinations of the perturbations
themselves. 
The problem is more involved when one is
interested in evaluating backreaction effects either at the classical or at
the quantum level.  
Such a problem has been faced in \cite{GMV1}, where a gauge
invariant prescription to average in cosmological setup was defined, and in
\cite{GMV2} where, using the previous prescription, a 
general-covariant and gauge invariant generalization of  the  effective 
equations (introduced in \cite{Buchert})
which describe the cosmological backreaction was given (see \cite{ULC} for a specific 
application of the prescription introduced in \cite{GMV1}).
The key point of our approach is to define  a general formula for the
classical or quantum average of any scalar quantity, on hypersurfaces on
which another given scalar quantity is constant. 
The physical object is the hypersurface defined by the above scalar, 
and not the gauge chosen to calculate such an average.

The gauge issue, in connection with the backreaction, has been already partially discussed in the literature (see e.g.  \cite{LS}-\cite{BBM}), in particular in
\cite{BBM} the authors, after introducing a generalization of the effective
equations presented in \cite{Buchert}, 
applied the effective equations to two different gauges.
Following \cite{GMV1,GMV2} the results of
\cite{BBM} should be seen in the following way. As already noted in 
\cite{GMV1} the standard average prescription present in the 
literature \cite{Buchert}, applied in a
particular gauge, can be seen as the gauge invariant prescription 
introduced in \cite{GMV1,GMV2} with
the average performed on the hypersurface defined by a scalar field
which is homogeneous in that particular gauge. 
So, from this point of view, it is clear that different
gauges should lead to different results for the standard averaging procedure,
since this correspond to averages performed on different hypersurfaces.
The different results obtained in \cite{BBM} actually correspond to the 
cosmological backreaction with respect to different observers/hypersurfaces.
A key question comes out: which 
hypersurface should be chosen to perform the average? The point is  
to understand what are the physical observers, with respect to whom
the backreaction should be calculated for the problem under consideration.

In this paper,  after  establishing a correspondence
between the standard average prescription and the one introduced in \cite{GMV1,GMV2},
 we will see how to define, for any given gauge,  an associated homogeneous scalar.
We will then show 
how the physical properties of the observers associated with such scalars are actually independent from the gauge chosen to perform the calculation. 
In particular we will present a gauge invariant description of  observers in geodesic motion. Such observers have been often suggested as the physically relevant ones (see, for 
example, \cite{LS}) for the calculation of the cosmological backreaction of 
the present space-time inhomogeneities. 
Furthermore we will discuss in detail the residual gauge dependence of the average 
prescription introduced in \cite{GMV1,GMV2}, when the average is performed on a 
finite volume (see Sect. 2 of \cite{GMV1}).  

The paper is organized as follows. 
In Sect.\ref{Sec2} we recall the averaging prescription introduced in \cite{GMV1,GMV2}.
In Sect.\ref{Sec3} we give some useful relations
among the scalars used to define the hypersurfaces where the 
average is performed. 
In particular the expression which defines the scalar that is homogeneous in a given 
gauge, and the relation which 
connects different homogeneous  scalars in different gauges.
Furthermore, we define the
geodesic motion of an observer, up to second order in perturbation
theory, in a perturbed FLRW space-time, and 
we introduce the hypersurface which defines such
observers in geodesic motion independently from the gauge considered.
In Sect.\ref{Sec4}
we study the residual gauge dependence of the averaging prescription as a function
of the size of the volume of integration.
Our conclusive remarks are presented in Sect.\ref{Sec5}.


\section{Averaging prescription: a recent proposal}
\label{Sec2}
\setcounter{equation}{0}

To evaluate the dynamical contribution of the inhomogeneities, present
in the metric and matter sectors, one has first to introduce a well defined averaging 
procedure with respect to the observer involved in the measurements.
For this purpose let us begin from the following four dimensional integral 
of a scalar $S(x)$ defined in \cite{GMV1}
\beq
F(S,\Om) = \int_{\Om(x)} d^{4} x  \sqrt{-g(x)} \,S(x) \equiv
 \int_{{\cal M}_4} d^{4} x  \sqrt{-g(x)} \,S(x)W_\Om(x)\,.
\label{1.1}
\eeq
The integration region $\Om \subseteq {\cal M}_4 $ is defined in terms of a suitable scalar window function
$W_\Om$, selecting a region with temporal boundaries determined by the space-like hypersurfaces  $\Sigma (A)$ over which a scalar field $A(x)$ takes constant values, and with spatial boundary determined by the coordinate condition $B<r_0$, where $B$ is a (positive) function of the coordinates with space-like gradient $\pa_\mu B$, and $r_0$ is a positive constant. As we are interested in the variation of the volume averages along the flow lines normal to $\Sigma (A)$, we choose the following window function \cite{GMV2}:
\beq
W_\Omega(x)=
u^{\mu} \nabla_\mu \theta(A(x)-A_0) 
\theta(r_0-B(x))
\label{Window}
\eeq
where $\theta$ is the Heaviside step function, and 
$u^{\mu}=-\frac{\partial^\mu A}{(-\partial^\tau A \partial^\nu A g_{\tau \nu})^{1/2}}$. 

Using the window function (\ref{Window}) we can define the averaging  of a scalar $S(x)$ as
\beq
\langle S \rangle_{A_0, r_0} = \frac{F(S, \Omega)}{F(1, \Omega)}= 
\frac{\int d^4x \sqrt{-g} \, S \,
u^{\mu} \nabla_\mu \theta(A(x)-A_0) 
\theta(r_0-B(x))}{\int  d^4x \sqrt{-g} \,
u^{\mu} \nabla_\mu \theta(A(x)-A_0) \,
\theta(r_0-B(x))}\,.
\label{1.4}
\eeq
In order to explicitly rewrite the above definition in terms of spatial (three-dimensional) integrals we can perform the covariant derivative and 
exploit the properties of the consequent delta function.
On considering the coordinate transformation  
to the bar coordinates, in which the proper time $t$ goes to $\tb$ defined by $t= h(\tb,x)$ and where the function $h$ is chosen to make the scalar field $A$ homogeneous, i.e.
\beq
A(h(\tb,x),x)= \overline A (\tb, x) \equiv A^{(0)}(\tb)\,,
\label{34}
\eeq
one can perform the time integration and obtain
\beq
\langle S  \rangle_{{A_0,r_0}}= 
\frac{\int_{\Sigma_{A_0}}  d^3x  \sqrt{|\overline{\gamma}(t_0, \vec{x})|} 
\,~ \overline{S}(t_0, \vec{x}) \,\theta(r_0-B(h(t_0, \vec{x}),
 \vec{x})}{\int_{\Sigma_{A_0}}
  d^3x  
\sqrt{|\overline{\gamma}(t_0, \vec{x})|} \,  
\theta(r_0-B(h(t_0, \vec{x}), \vec{x}))}
\label{1.5}
\eeq
where we have called $t_0$ the time $\bar{t}$ when $A^{(0)}(\bar{t})$ takes the
constant values $A_0$ and we are averaging on a section of the 
three-dimensional hypersurface $\Sigma_{A_0}$,  hypersurface where $A(x)=A_0$.
In this notation, for example, $\bar{S}$ is the variable $S$ transformed to the 
coordinate frame in which $A(x)$ is homogeneous.
In this framework the hypersurface $A(x)=A_0$ is the one which identifies our observers (see Sect. 3). 

As discussed in \cite{GMV1},
if $B$ is not a scalar, the spatial boundary can be a source of
breaking of  covariance and gauge invariance.
In the limit of large averaging volume with respect to the typical scale of the 
inhomogeneities, all the gauge dependence disappears (\cite{GMV1}) and the classical 
average, or the correspondent vacuum expectation value (see \cite{GMV1}), is 
uniquely characterized by the hypersurface of integration $A(x)=A_0$ 
regardless from the gauge chosen to perform the calculation.
In Sect. \ref{Sec3} we will give some useful tool to define the scalar $A(x)$ and,
using a particular example, we will see as the physical properties    
of the observers associated to such scalar are independent from the gauge chosen,
while in Sect.\ref{Sec4} we will discuss in more quantitative terms such a breaking.


\section{Gauges vs observers in a gauge invariant averaging prescription}
\label{Sec3}
\setcounter{equation}{0}

As stated in \cite{GMV1,GMV2} and in the introduction, 
backreaction effects are usually
calculated, in the literature, in different 
gauges using the standard averaging prescription
\cite{Buchert}. 
In this way the results can be seen as  
particular cases of the  average prescription recalled
in Sect.\ref{Sec2} only if the 
gauge considered is the one where the scalar $A(x)$, which defines the
hypersurface on which we perform the average, is homogeneous.
In such a gauge, using the ADM formalism, the standard averaging prescription 
\cite{Buchert} is also covariantly defined. In \cite{Buchert} this corresponds 
to identify $A(x)$ with the scalar potential of the fluid and choose the so-called 
covariant fluid gauge \cite{EB}.

To better understand the connection above let us see in detail how it is always 
possible to define a 
particular scalar which is homogeneous in a particular given gauge.

\subsection{Scalars/Gauges correspondence}

We shall work in the context of a spatially flat, FLRW background
geometry, and we expand our background fields $\{g_{\mu\nu}\}$
up to second order in the non-homogeneous perturbations, without fixing
any gauge, as follows:
\bea
& & g_{00}= -1-2 \a^{(1)}-2 \a^{(2)}, ~~~~~~~\,\,\,\,\,\,\,\,\,\,
g_{i0}=-{a\over2}\left(\beta^{(1)}_{,i}+B^{(1)}_i\right) 
-{a\over2}\left(\beta^{(2)}_{,i}+B^{(2)}_i\right)\,,
\nonumber \\
& & 
g_{ij} = a^2 \left[ \delta_{ij} \left(1-2 \psi^{(1)}-2 \psi^{(2)}\right)
+D_{ij} (E^{(1)}+E^{(2)}) +
{1\over 2} \left(\chi^{(1)}_{i,j}+\chi^{(1)}_{j,i}+h^{(1)}_{ij}\right) \right.
\nonumber \\
& & \left.
\,\,\,\,\,\,\,\,\,\,\,\,\,\,+
{1\over 2} \left(\chi^{(2)}_{i,j}+\chi^{(2)}_{j,i}+h^{(2)}_{ij}\right)\right],
\label{FLRWpGG}
\end{eqnarray}
where $D_{ij}=\partial_i \partial_j- \delta_{ij} (\nabla^2/3)$, and
$a=a(t)$ is the scale factor of the homogeneous FLRW  metric. Here 
 $\a^{(1)}$, $\b^{(1)}$, $\psi^{(1)}$, $E^{(1)}$ are pure scalar 
first-order perturbations, 
$B^{(1)}_i$ and $\chi^{(1)}_i$ are transverse vectors ($\pa^i B^{(1)}_i=0$ and 
$\pa^i \chi^{(1)}_i=0$), $h^{(1)}_{ij}$ is a traceless and transverse tensor 
($ \pa^i h^{(1)}_{ij}=0=h^{(1)\,i}_i$), and the same notation applies to
 the 
case of the second-order perturbations.
 
From Eq.(\ref{FLRWpGG}) one obtains 10 degrees of freedom (which must be added 
to those coming from the matter sector) which are in part redundant.
To obtain a set of well defined equations (Einstein equations + equations 
of motion of the matter sector),  order by order, we have, for example,
to set to zero two scalar perturbations and one vector perturbation.

The choice of such variables is called a ``gauge'' choice  and can be 
performed using the gauge transformation, associated to a  ``infinitesimal'' 
coordinate transformation, from the general gauge (\ref{FLRWpGG}) to the 
particular gauge chosen. 
Such ``infinitesimal'' 
coordinate transformation is parametrized by the  first-order, $\ep_{(1)}^\mu$, 
and second-order, $\ep_{(2)}^\mu$,  generators as \cite{BMMS}
\beq
x^\mu \rightarrow \tilde{x}^\mu= x^\mu + \epsilon^\mu_{(1)} +\frac{1}{2}
\left(\epsilon^{\nu}_{(1)}\pa_\nu \epsilon^{\mu}_{(1)} + 
\epsilon^{\mu}_{(2)}\right) + \dots
\label{311}
\eeq
where
\beq
\ep_{(1)}^\mu= \left( \ep_{(1)}^0, \hat{\ep}_{(1)}^{\,i} 
\right), ~~~~~~~~~
\ep_{(2)}^\mu= \left( \ep_{(2)}^0, \hat{\ep}_{(2)}^{\,i} \right) 
\label{312}
\eeq
and we can explicitly separate the scalar and the pure transverse
vector part of $\hat{\ep}_{(1)}^i$, $\hat{\ep}_{(2)}^i$ as 
\be 
\hat{\ep}_{(1)}^{\,i}=\pa^i \ep_{(1)}+ \ep_{(1)}^i \,\,\,\,\,\,\,\,\,\,,
\,\,\,\,\,\,\,\,\,
\hat{\ep}_{(2)}^{\,i}=\pa^i \ep_{(2)}+ \ep_{(2)}^i\,.
\ee 
Under the associated gauge transformation 
(or local field reparametrization) - where, by definition, old
and new fields are evaluated at the same space-time point x - 
 a general tensor changes, to first and second 
order, as
\be
T^{(1)} \rightarrow \tilde{T}^{(1)}=  T^{(1)}-L_{\ep_{(1)}} T^{(0)}
\label{GGT1}
\ee
\be
T^{(2)} \rightarrow \tilde{T}^{(2)}=  T^{(2)}-L_{\ep_{(1)}} T^{(1)}
+\frac{1}{2}\left( L^2_{\ep_{(1)}} T^{(0)}-L_{\ep_{(2)}} T^{(0)}\right)
\label{GGT2}
\ee
where $L_{\ep_{(1,2)}}$ is the Lie derivative respect the vector 
$\ep^\mu_{(1,2)}$.
In particular the scalar metric perturbations change,  to first order, as
\be
\alphat^{(1)} = \alpha^{(1)} - \dot \epsilon^0_{(1)},
\label{415}
\ee
\be
\betat^{(1)} = \beta^{(1)} - \frac{2}{a} \epsilon^0_{(1)} +2 a \dot \epsilon_{(1)},
\label{416}
\ee
\be
\tilde{\psi}^{(1)} =\psi^{(1)}+ H \epsilon^0_{(1)}+\frac{1}{3} 
\nabla^2\epsilon_{(1)},
\label{417}
\ee
\be
\tilde{E}^{(1)}=E^{(1)} -2\epsilon_{(1)} \,.
\label{418}
\ee
and to second order, as
\bea
\alphat^{(2)} &=& \alpha^{(2)}-\frac{1}{2}\dot \epsilon^0_{(2)}
- \epsilon^0_{(1)} \dot{\alpha}^{(1)}-2 \alpha^{(1)}
\dot \epsilon^0_{(1)}+\frac{1}{2}
\epsilon^0_{(1)}\ddot \epsilon^0_{(1)} +\left(\dot 
\epsilon^0_{(1)}\right)^2-\hat{\ep}_{(1)}^{\,i} \alpha_{,i}^{(1)}
\nonumber \\ & & 
+\frac{1}{2}
\left(\hat{\ep}_{(1)}^{\,i} \epsilon^0_{(1),i}\right)^.
-\frac{a^2}{2}
\dot{\hat{\ep}}_{(1)}^{\,i}\dot{\hat{\ep}}_{(1)i}-\frac{a}{2}\left(\beta^{(1)}_{,i}+
B_i^{(1)}\right) 
\dot{\hat{\ep}}_{(1)}^{\,i},
\label{422}
\\
\tilde{\beta}^{(2)}&=& \beta^{(2)} -\frac{1}{a} \epsilon_{(2)}^0
+ a \dot{\epsilon}_{(2)} +\frac{1}{2a}\frac{d}{d t} (\epsilon_{(1)}^0)^2
+\frac{1}{a}\hat{\ep}_{(1)}^{\,i}
\epsilon^0_{(1),i}- \hat{\ep}_{(1)}^{\,i}\beta^{(1)}_{,i}
+\frac{\partial^i}{\nabla^2} \Upsilon_i\,,
\label{423}\\
\tilde{\psi}^{(2)} &=& \psi^{(2)}+
\frac{H}{2}\epsilon^0_{(2)}+\frac{1}{6}\nabla^2 \epsilon_{(2)}
-\epsilon_{(1)}^0 \left(2 H \psi^{(1)}+\dot{\psi}^{(1)}
\right) -
\frac{H}{2} \epsilon^0_{(1)} \dot \epsilon^0_{(1)} 
\nonumber \\
& &
-\frac{\epsilon^{0 \, 2}_{(1)}}{2} \left( \dot H + 2 H^2 \right) 
-\frac{H}{2}\epsilon^{0}_{(1),i}\hat{\ep}^{\,i}_{(1)}
-\psi_{,i}^{(1)} \hat{\ep}^{\,i}_{(1)}
-\frac{1}{6} \Pi^i_i\,,
\label{424}
\\
\tilde{E}^{(2)} &=& E^{(2)}- \epsilon_{(2)}
-\frac{1}{2}\frac{1}{\nabla^2}\Pi^i_i
+\frac{3}{2}\frac{\partial^i\partial^j}{(\nabla^2)^2}\Pi_{ij}\,,
\label{425}
\end{eqnarray}
where
\begin{eqnarray}
\Upsilon_i &=& \frac{2}{a} \epsilon_{(1),i}^0 \dot{\epsilon}_{(1)}^0
- \epsilon_{(1)}^0  (\dot{\beta}^{(1)}_{,i}+\dot{B}^{(1)}_i)-
H \epsilon_{(1)}^0 (\beta^{(1)}_{,i}+B_i^{(1)}) -\dot{\epsilon}_{(1)}^0 
(\beta^{(1)}_{,i}+B_i^{(1)})  
\nonumber \\ & &
- \frac{4}{a} \alpha^{(1)} \epsilon_{(1),i}^0 
-2a \dot{\hat{\ep}}_{(1)i} \left( 2 H \epsilon^0_{(1)}+\frac{1}{2}
\dot{\epsilon}^0_{(1)}+2 \psi^{(1)}\right)-a \ddot{\hat{\ep}}_{(1)i}\epsilon^0_{(1)}
\nonumber \\ & &
-a\left(\hat{\ep}^{\,j}_{(1)}\dot{\hat{\ep}}_{(1)i,j}+2 
\hat{\ep}^{\,j}_{(1),i}\dot{\hat{\ep}}_{(1)j}
+\dot{\hat{\ep}}^{\,j}_{(1)}
\hat{\ep}_{(1)i,j} \right)- \hat{\ep}^{\,j}_{(1)}B_{i,j}^{(1)}
-\hat{\ep}^{\,j}_{(1),i}B_{j}^{(1)}
\nonumber \\ & &
-2 a \dot{\hat{\ep}}^{\,j}_{(1)} 
\left[-D_{ij} E^{(1)}-\frac{1}{2}\left(\chi_{i,j}^{(1)}+\chi_{j,i}^{(1)}-
h_{ij}^{(1)}\right)\right]
\end{eqnarray}
\begin{eqnarray}
\Pi_{i j}&=&-2 \epsilon^{0}_{(1)} D_{i j} \left(H E^{(1)}
+\frac{\dot{E}^{(1)}}{2}\right) 
-\hat{\ep}^{\,k}_{(1)} D_{i j} E_{,k}^{(1)}
-\hat{\ep}^{\,k}_{(1),j} D_{i k} E^{(1)}
-\hat{\ep}^{\,k}_{(1),i} D_{j k} E^{(1)}
\nonumber \\
& & 
+2 H \epsilon^{0}_{(1)} \left(\hat{\ep}_{(1)i,j}+\hat{\ep}_{(1)j,i}\right)
+2 \psi^{(1)} \left(\hat{\ep}_{(1)i,j}+\hat{\ep}_{(1)j,i}\right) 
-\epsilon^{0}_{(1)}\left[H \left( 
\chi_{i,j}^{(1)}+\chi_{j,i}^{(1)}+h_{ij}^{(1)}\right)
\right. \nonumber \\
& & \left.
+\frac{1}{2} \left(\dot{\chi}_{i,j}^{(1)}+\dot{\chi}_{j,i}^{(1)}
+\dot{h}_{ij}^{(1)}\right)\right]
+\frac{1}{2}\epsilon^{0}_{(1)}\left(\dot{\hat{\ep}}_{(1)i,j}
+\dot{\hat{\ep}}_{(1)j,i}\right)
-\frac{1}{a^2} \epsilon^{0}_{(1),i}\epsilon^{0}_{(1),j}
\nonumber \\
& & 
+\hat{\ep}_{(1)k,i}\hat{\ep}^{\,k}_{(1),j}
+\frac{1}{2}\hat{\ep}^{\,k}_{(1)}
\left(\hat{\ep}_{(1)i,j k}+\hat{\ep}_{(1)j,i k}\right)
-\frac{1}{2}\hat{\ep}^{\,k}_{(1)} \left(
\chi_{i,j k}^{(1)}+\chi_{j,i k}^{(1)}+
h_{i j,k}^{(1)}\right)
\nonumber \\
& & 
+\frac{1}{2}\left(\dot{\hat{\ep}}_{(1) i}\epsilon^{0}_{(1),j}+
\dot{\hat{\ep}}_{(1) j}\epsilon^{0}_{(1),i}\right)
+\frac{1}{2 a}\left(\beta_{,i}^{(1)}+B_i^{(1)}\right) \epsilon^{0}_{(1),j}
+\frac{1}{2 a}\left(\beta_{,j}^{(1)}+B_j^{(1)}\right) \epsilon^{0}_{(1),i}
\nonumber \\
& & 
+\frac{1}{2}\left(\hat{\ep}_{(1)j, k}\hat{\ep}^{\,k}_{(1),i}+
\hat{\ep}_{(1)i, k}\hat{\ep}^{\,k}_{(1),j}\right)
-\frac{1}{2}\left(\chi_{i,k}^{(1)}+\chi_{k,i}^{(1)}+h_{i k}^{(1)}\right) \hat{\ep}^{\,k}_{(1),j}
\nonumber \\
& & 
-\frac{1}{2}\left(\chi_{j,k}^{(1)}+\chi_{k,j}^{(1)}+h_{j k}^{(1)}\right) \hat{\ep}^{\,k}_{(1),i}\,.
\end{eqnarray}

For the vector metric perturbations one obtains, to first order
\be
\tilde{B}_i^{(1)}=B_i^{(1)}+2 a \dot{\ep}_{(1)i}
\ee
\be 
\tilde{\chi}_i^{(1)}=\chi_i^{(1)}-2 \ep_{(1)i}
\ee
and to second order \cite{MMMW}
\be
\tilde{B}^{(2)}_i=B^{(2)}_i+a \dot{\ep}_{(2)i}-\frac{\partial_i \partial^j}{\nabla^2}
\Upsilon_j +\Upsilon_i
\ee
\be 
\tilde{\chi}^{(2)}_i=\chi^{(2)}_i-\ep_{(2)i}
-2 \frac{\partial_i\partial^j\partial^k}{(\nabla^2)^2}\Pi_{j k}
+2 \frac{\partial^j}{\nabla^2}\Pi_{i j}
\ee
The metric tensor perturbation is gauge invariant to first order
($\tilde{h}_{i j}^{(1)}=h_{i j}^{(1)}$)  
while to second order one obtains \cite{MMMW}
\be
\tilde{h}_{i j}^{(2)}=h_{i j}^{(2)}+2 \Pi_{i j}
+\left(\frac{\partial_i \partial_j}{\nabla^2}-\delta_{i j}\right) \Pi^k_k
+\left(\frac{\partial_i \partial_j}{\nabla^2}+\delta_{i j}\right) 
\frac{\partial^k \partial^l}{\nabla^2}\Pi_{k l}
-\frac{2}{\nabla^2}\left(\partial_i \partial^k \Pi_{j k}
+\partial_j \partial^k \Pi_{i k}\right)\,.
\ee
Let us note that the differences between Eqs.(\ref{417},\ref{418})-(\ref{424},\ref{425}) and corresponding ones of \cite{MMMW} are connected with a slightly different definition of the scalar 
perturbations associated with $g_{ij}$.

To conclude the associated gauge transformation of a scalar 
field $A(x)=A^{(0)}(t)+A^{(1)}(x)+A^{(2)}(x)$
is, to first order, 
\be
A^{(1)}~~ \rightarrow ~~\ti{A}^{(1)}=A^{(1)}-\ep_{(1)}^0 \dot{A}^{(0)},
\label{315}
\ee
and, to second order, 
\bea
A^{(2)} ~~\rightarrow &&~~\ti{A}^{(2)}=  A^{(2)}-\ep_{(1)}^0 \dot{A}^{(1)}
-\left(\ep_{(1)}^i+\partial^i \ep_{(1)}\right) \partial_i A^{(1)}
\nonumber \\ 
&&+\frac{1}{2}\left[\ep_{(1)}^0 
\partial_t (\ep_{(1)}^0 \dot{A}^{(0)})+ \left(\ep_{(1)}^i+\partial^i \ep_{(1)}\right)\partial_i
\ep_{(1)}^0 \dot{A}^{(0)}-\ep_{(2)}^0 \dot{A}^{(0)}\right]\,.
\label{316}
\eea 

In this way the general scalar $A(x)$ homogeneus in the gauge 
chosen
(identified by the generators $\ep_{(1)}^\mu$ and $\ep_{(2)}^\mu$)
will be given by
\begin{eqnarray}
A^{(0)}(t) &=&A^{(0)}(t) \nonumber \\
A^{(1)}(x)&=&\dot{A}^{(0)} \epsilon_{(1)}^0
\nonumber \\
A^{(2)}(x)&=&\frac{\dot{A}^{(0)}}{2} \epsilon_{(1)}^0\dot{\epsilon}_{(1)}^0+
\frac{\ddot{A}^{(0)}}{2} \epsilon_{(1)}^{0\,2}+\frac{\dot{A}^{(0)}}{2}\left[
\left(\epsilon_{(1)}^i+\partial^i \epsilon_{(1)}\right)
\partial_i \epsilon_{(1)}^0+ \epsilon_{(2)}^0\right]
\label{Scalar_General}
\end{eqnarray}
we have a class of scalars defined by their homogeneous value $A^{(0)}(t)$ 
which can take any value.
From another point of view, the coordinate
transformation, that goes from a general gauge to the bar gauge where
$A(x)$ is homogeneous, fixes the
corresponding transformation parameters as follows:
\beq
\ep_{(1)}^0=\frac{A^{(1)}}{\dot{A}^{(0)}}, 
~~~~~~
\ep_{(2)}^0=2 \frac{A^{(2)}}{\dot{A}^{(0)}}- \frac{A^{(1)}
\dot{A^{(1)}}}{\left(\dot{A}^{(0)}\right)^2} 
- \frac{1}{\dot{A}^{(0)}}  \left(\pa^i \ep_{(1)}+ \ep_{(1)}^i\right) 
\partial_i  A^{(1)},
~~~~~~
\label{41}
\eeq
as it is easy to control the two set of equations 
(\ref{Scalar_General}) and (\ref{41}) are one the inverse of the other.

To check the results in Eqs. (\ref{Scalar_General}) one can consider the simple case of a space-time sourced by a single scalar field 
$\phi(x)=\phi^{(0)}(t)+\varphi^{(1)}(x)+\varphi^{(2)}(x)$,
and choose the uniform field gauge (UFG), namely the gauge for which such
scalar $\phi(x)$ is homogeneous, 
as the gauge where our scalar $A(x)$ should be homogeneous.
In such a case is pretty obvious that $A(x)$ is given by a general function of 
$\phi(x)$, namely $A=A(\phi(x))$.
If we now expand up to second order in perturbation theory we obtain
\be
A(x)=A(\phi(x))=A^{(0)}+\frac{\dot{A}^{(0)}}{\dot{\phi}^{(0)}} \varphi^{(1)} 
+\frac{\dot{A}^{(0)}}{\dot{\phi}^{(0)}} \varphi^{(2)}
+\frac{\dot{A}^{(0)}}{2 \dot{\phi}^{(0)\,2}} \left(
\frac{\ddot{A}^{(0)}}{\dot{A}^{(0)}}-
\frac{\ddot{\phi}^{(0)}}{\dot{\phi}^{(0)}}\right)\varphi^{(1)\,2} \,,
\label{42}
\ee
which is, in fact, the expected solution  of Eqs. (\ref{Scalar_General}).

As pointed out the scalar $A(x)$ is used to identify our observers. 
They are the ones sitting on the spacelike hypersurfaces over which the scalar field 
$A(x)$ takes constant values, and their motion 
is given by the comoving 4-velocity
\be
u_\mu= - \frac{\partial_\mu A}{\left(-\partial_\sigma A \partial^\sigma 
A\right)^{1/2}}
\label{Comoving_velocity}
\ee
which takes the general form  
\begin{eqnarray}
u_\mu^{(0)} \!\!\!\!&=&\!\!\!\! \left(-1, \vec{0} \right) \nonumber \\
u_\mu^{(1)} \!\!\!\!&=&\!\!\!\! \left(-\alpha^{(1)}, -\epsilon_{(1) , i}^0 \right)
\nonumber \\ 
u_\mu^{(2)} \!\!\!\!&=&\!\!\!\!  \left(\frac{\alpha^{(1)\, 2}}{2}
\!+\!\frac{\epsilon^{0}_{(1) ,j}}{2 a}
\left(\beta^{(1),j}+B^{(1)\,j}\right)\!-\!\frac{1}{2 a^2} 
\epsilon^{0}_{(1) ,j}\epsilon^{0 , j}_{(1)}\!
-\!\frac{1}{8}
\left(\beta^{(1)}_{,j}\!+\!B^{(1)}_j\right) \left(\beta^{(1),j}\!+\!B^{(1)\,j}
\right)\!-\!\alpha^{(2)},                    
\right.
\nonumber \\
& & \left. \!\!\!\!\!\!
\frac{1}{2} \epsilon^{0}_{(1) ,i }\dot{\epsilon}^{0}_{(1)}
- \frac{1}{2} \epsilon^{0}_{(1)}\dot{\epsilon}^{0}_{(1) ,i}
- \alpha^{(1)} \epsilon^{0}_{(1) ,i}
-\frac{\epsilon^0_{(2) , i}}{2}
-\frac{1}{2}\left(\hat{\epsilon}^{j}_{(1) ,i}\epsilon^{0}_{(1) ,j}+
\hat{\epsilon}^{j}_{(1)}\epsilon^{0}_{(1) ,j i} \right)\right)
\label{CGeneral_2}
\end{eqnarray}

This normal vector to the hypersurface is the one which describes the physical properties of our 
observers. As we will see in a particular example in the next subsection, such 
properties are the same in all  possible gauges, the vector 
describes a gauge invariant set of properties. 
Let us also put in evidence that  such a vector is independent from 
the homogeneous value 
$A^{(0)}(t)$ of the corresponding scalar. In general, this value can be
chosen at will
and does not influence the physical properties of the observers
 and the physical meaning of our averages.
For example, we can choose $A^{(0)}(t)=t$ to have standard derivatives in the
gauge invariant formulation of the cosmological backreaction \cite{GMV2}.

An interesting result is that all these particular scalars, homogeneous in
different gauges, can be connected to each other by a simple formula. 
Let us call with $A_i$ and $A_j$ the general expressions, without fixing any gauge,
of the scalars homogeneous in the gauge $i$ and $j$, one obtains the following
result
\be
A_i=A_j-\left(A^{(1)}_j\right)_i-\left(A^{(2)}_j\right)_i 
-\left(\epsilon_{(1)}^\mu\right)_i \partial_\mu (A^{(1)}_j)_i
\label{43}
\ee 
where $(A_j^{(n)})_i$ is the value that $A_j^{(n)}$ takes after a gauge
transformation from the general gauge to the gauge $i$ (so
it is gauge invariant for construction), and 
$\left(\epsilon_{(1)}^\mu\right)_i$ is the value that we have to give to
the parameter $\epsilon_{(1)}^\mu$ to go from a general gauge to the
particular gauge $i$.
It is easy to understand that this last quantity changes in the following
way under a gauge transformation identified by a generator $\epsilon_{(1)}^\mu$
\be
\left(\epsilon_{(1)}^\mu\right)_i \rightarrow 
\left(\epsilon_{(1)}^\mu\right)_i-\epsilon_{(1)}^\mu
\label{44}
\ee
so the quantity obtained by using Eq.(\ref{43}) is homogeneous in the
gauge $i$ and changes as a scalar up to second order. Starting with a
scalar homogeneous in a given gauge we can easily obtain a scalar homogeneous
in any other given gauge using Eq.(\ref{43}). 
This is a general formula that can be
generalized to any order.

To conclude we have that any given gauge has a class of scalar fields,
homogeneous in that gauge, while for any given scalar there is a class
of bar gauges (the parameters of the coordinate transformation which gives
the bar gauge are only partially determined by $A(x)$, see Eq.(\ref{41})).


\subsection {A privileged observer?}

One of the central problems in the
calculation of  backreaction effects from inhomogeneities 
is to choose the hypersurface with respect to which our
observers are defined. 
Such a choice will obviously depend on the particular backreaction problem 
considered.
In the particular case of the backreaction of large scale structure on the present 
cosmological evolution different authors (see, for example, \cite{LS})
have claimed that such physical observers are given by the
observers in geodesic motion. 
This property should be a gauge independent property of our observers, i.e.
the results of the backreaction should
depend only on the hypersurface which defines the observers and not on 
the gauge chosen to perform the calculation.

An observer in geodesic motion is described by a velocity vector $v_\mu$ 
(normalized as $v_\mu v^\mu=-1$) which
satisfies the following equation
\be
g^{\sigma \tau} v_\sigma \nabla_\tau v_\mu=0 \,,
\label{General_GE}
\ee
using the perturbed FLRW metric given in Eq.(\ref{FLRWpGG}) we obtain
the following result up to second order in perturbation theory
\begin{eqnarray}
v_\mu^{(0)} &=& \left(-1, \vec{0} \right) \nonumber \\
v_\mu^{(1)} &=& \left(-\alpha^{(1)}, -\int dt \alpha^{(1)}_{,i} \right) 
\nonumber \\
v_\mu^{(2)} &=& \left(v_0^{(2)}, \int dt v^{(2)}_{0,i} \right)\,, 
\label{GE_2}
\end{eqnarray}
where
\begin{eqnarray}
v_0^{(2)} &=&  \frac{\alpha^{(1)\, 2}}{2}-\frac{1}{2 a^2}
\left(\int dt \alpha^{(1)}_{,j}\right) \left(\int dt \alpha^{(1),j} 
\right)+ \frac{1}{2 a}
\left(\int dt
 \alpha^{(1)}_{,j}\right)\left(\beta^{(1),j}+B^{(1)\,j}\right)
\nonumber \\
& & 
-\frac{1}{8}
\left(\beta^{(1)}_{,j}+B^{(1)}_j\right) \left(\beta^{(1),j}+B^{(1)\,j}
\right)                     
-\alpha^{(2)}
\end{eqnarray}
Let us note that this vector $v_\mu$ will take the standard zero order
form $(-1, \vec{0})$ 
in the so-called synchronous gauge (where 
$\alpha^{(1)}=\alpha^{(2)}=0$, $\beta^{(1)}=0$ and
$B_i^{(1)}=0$).
 This  suggests that the scalar
which defines the ``right'' hypersurface should be the one which is
homogeneous in such a synchronous gauge. 
Such scalar field $A(x)$ can be constructed starting from 
Eqs.(\ref{Scalar_General}) or using the 
relation in (\ref{43}) and the scalar homogeneus in the UFG given in (\ref{42}).
After some calculation 
we obtain that this is given by
\begin{eqnarray}
A^{(0)}(t)&=& A^{(0)}(t) \nonumber \\
A^{(1)}(x)&=& \dot{A}^{(0)} \int dt \alpha^{(1)} \nonumber \\
A^{(2)}(x)&=& \frac{1}{2} \ddot{A}^{(0)} \left(\int dt \alpha^{(1)}\right)^2
+\dot{A}^{(0)} \int dt \left[-\frac{1}{2} \alpha^{(1)\,2}
+\frac{1}{2 a^2} \left(\int dt \alpha^{(1)}_{,j}\right) 
\left(\int dt \alpha^{(1),j}\right)
\right. \nonumber \\
& & \left. \!\!\!\!\!\!\!\!\!\!\!\!\!\!\!\!\!\!- \frac{1}{2 a}
\left(\int dt \alpha^{(1)}_{,j}\right)\left(\beta^{(1),j}+B^{(1)\,j}\right)
+\frac{1}{8}
\left(\beta^{(1)}_{,j}+B^{(1)}_j\right)
\left(\beta^{(1),j}+B^{(1)\,j}\right)  
+\alpha^{(2)}\right]\,
\label{A_SG}
\end{eqnarray}
and we can easily check that
\be
u_\mu \equiv v_\mu~,
\ee
namely the observers defined by this scalar are always in
geodesic motion independently from the gauge chosen.
This is a gauge independent property namely a physical property of the 
observers.


\section{Gauge dependence for finite averaging volume}
\label{Sec4}
\setcounter{equation}{0} 

In the calculation of the backreaction effects the region of integration has 
a typical size that goes from tens of Mpc to the Hubble radius.
Consequently,  it is important to investigate how the average  introduced in \cite{GMV1,GMV2}  depends on the
gauge chosen as a consequence of the finiteness of the 
region of integration.
The gauge dependence, up to second order in the cosmological perturbations, of a 4-dimensional integral, using the window function 
(\ref{Window}) and the coordinate trasformation (\ref{311}), is given by (see, also, \cite{GMV1})
\bea
& &\ti{F}(\ti{S}, \Om) - {F}({S}, \Om)= - \int d^{4}x \sqrt{-g} \, S(x) u^\mu \nabla_\mu \theta(A(x)- A_{0}) \, \delta\left(r_0-B(x)\right)
\nonumber \\ & & \,\,\,\,\,\,\,\,\,\,\,\,
\left\{\frac{\partial B(x)}{\partial
x^\mu}\epsilon^\mu_{(1)}+\frac{1}{2}\frac{\partial B(x)}{\partial
x^\mu}\left[\epsilon^\nu_{(1)}\partial_\nu \epsilon^\mu_{(1)}+\epsilon^\mu_{(2)}\right]
+\frac{1}{2}\frac{\partial^2 B(x)}{\partial x^\mu \partial x^\nu }
\epsilon^\mu_{(1)}\epsilon^\nu_{(1)}\right\}
\label{4.1}
\eea
where a $\tilde{}$ indicates  gauge trasformed quantities.
Using this result in the  averaging prescription defined in (\ref{1.4})
one obtains:
\bea
\langle \tilde{S} \rangle_{A_0, r_0} &=& \frac{\tilde{F}(\tilde{S}, 
\Omega)}{\tilde{F}(1, \Omega)}= \left\{F(S, \Omega) - 
\int d^{4}x \sqrt{-g} \, S(x) u^\mu \nabla_\mu \theta(A(x)- A_{0}) 
\delta\left(r_0-B(x)\right)\right. \nonumber \\
& & \left. 
\left[ \frac{\partial B(x)}{\partial
x^\mu}\epsilon^\mu_{(1)}+\frac{1}{2}\frac{\partial B(x)}{\partial
x^\mu}\left[\epsilon^\nu_{(1)}\partial_\nu \epsilon^\mu_{(1)}+\epsilon^\mu_{(2)}\right]
+\frac{1}{2}\frac{\partial^2 B(x)}{\partial x^\mu \partial x^\nu }
\epsilon^\mu_{(1)}\epsilon^\nu_{(1)}   \right] \right\} \nonumber \\
& &  \left\{F(1, \Omega) - 
\int d^{4}x \sqrt{-g} \, u^\mu \nabla_\mu \theta(A(x)- A_{0}) 
\delta\left(r_0-B(x)\right)\right. \nonumber \\
& & \left. 
\left[ \frac{\partial B(x)}{\partial
x^\mu}\epsilon^\mu_{(1)}+\frac{1}{2}\frac{\partial B(x)}{\partial
x^\mu}\left[\epsilon^\nu_{(1)}\partial_\nu \epsilon^\mu_{(1)}+\epsilon^\mu_{(2)}\right]
+\frac{1}{2}\frac{\partial^2 B(x)}{\partial x^\mu \partial x^\nu }
\epsilon^\mu_{(1)}\epsilon^\nu_{(1)}   \right] \right\}^{-1}\,.
\label{2.5}
\eea
 Keeping only  terms up to second order in the
cosmological perturbations, one obtains, after some algebra, the following 
result for the gauge dependence:
\begin{eqnarray}
\langle \tilde{S} \rangle_{A_0, r_0}-\langle S \rangle_{A_0, r_0}&=&
\frac{\int_{\Sigma_{A_0}} d^3x  
\,~ \overline{S}^{(1)} 
\, \theta(r_0-B(t_0, \vec{x}))}{\int_{\Sigma_{A_0}} d^3x  
\, \theta(r_0-B(t_0, \vec{x}))} 
\frac{\int_{\Sigma_{A_0}} d^3x  
\,~ \frac{\partial B}{\partial x^\mu} 
\bar{\epsilon}^\mu_{(1)} \, 
\delta(r_0-B(t_0, \vec{x}))}{\int_{\Sigma_{A_0}} d^3x  
\, \theta(r_0-B(t_0, \vec{x}))} \nonumber \\
& & - \frac{\int_{\Sigma_{A_0}} d^3x  
\,~ \overline{S}^{(1)}  \frac{\partial B}{\partial x^\mu}  
\bar{\epsilon}^\mu_{(1)} \, \delta(r_0-B(t_0,
\vec{x}))}{\int_{\Sigma_{A_0}} 
d^3x  \, \theta(r_0-B(t_0, \vec{x}))}
\label{4.4}
\end{eqnarray}
where $\bar{\epsilon}^\mu_{(1)}$ is the first order vector generator of
the gauge transformation of a bar quantity. Namely, if we consider a scalar
and vector quantity in the bar coordinate and we apply a gauge
transformation we obtain new quantities that can be
obtained by the general definition of a gauge transformation with 
 $\epsilon^\mu_{(1)}$  substituted by $\bar{\epsilon}^\mu_{(1)}$.
In practice we obtain that
\beq
\bar{\epsilon}^\mu_{(1)}=(0,\hat{\epsilon}^i_{(1)})\,.
\label{bar_epsilon_tot}
\ee
Let us remark some points about this result: first, the result is independent
from the window function used to carry out the classical average. If we repeat the
procedure starting from $W_\Omega(x)=\delta(A(x)-A_0)\theta(r_0-B(x))$ 
instead of $u^{\mu} \nabla_\mu \theta(A(x)-A_0) \theta(r_0-B(x))$
we obtain the same result.
In general, we obtain the result in Eq.(\ref{4.4}) starting from a window function 
$\delta(A(x)-A_0)\theta(r_0-B(x)) C(x)$ 
with $C(x)$ any scalar field with a non-vanishing homogeneous value. 
Second, the gauge dependence goes to zero 
for $r_0 \rightarrow +\infty $.
Third, regardless of the scalar $S$ the gauge dependence 
is always at least of  second order, so 
this will be always subleading if in the average 
$\langle S \rangle_{A_0, r_0}$ there is a non zero first order contribution.
And, in the end, if $\bar{S}^{(1)}=0$ the average is gauge invariant 
up to second order.


\subsection{An example}

Let us now consider a simple example where we take a sphere as spatial 
boundary. Namely we take $B(t_0,\vec{x})=\left(x^2+y^2+z^2\right)^{1/2}
=|\vec{x}|$.
Having set the spatial boundary our result will depend only on the structure of 
$\bar{S}^{(1)}$ and $\hat{\epsilon}_{(1)}^i$.
In general, if we consider a cosmological background sourced by an
energy-momentum tensor with no transverse part for $T^0_i$, namely with
no term $\hat{T}^0_i$ such that $\partial^i\hat{T}^0_i=0$, we have
that the first order vector perturbations can be set to zero \cite{MMT} 
regardless of the gauge and
we can set, from Eq.(\ref{bar_epsilon_tot}), 
$\bar{\epsilon}^\mu_{(1)}=(0,\partial^i \epsilon_{(1)})$.

Using these assumptions Eq.(\ref{4.4}) can be
simplified to
\begin{eqnarray}
\langle \tilde{S} \rangle_{A_0, r_0}-\langle S \rangle_{A_0, r_0}&=&
\frac{\int d^3x  
\,~ \overline{S}^{(1)} 
\, \theta(r_0-|\vec{x}|)}{\frac{4}{3}\pi r_0^3} 
\frac{\int d^3x  
\,~ \vec{\nabla} |\vec{x}| 
\cdot 
\vec{\nabla} \epsilon_{(1)}
 \, \delta(r_0-|\vec{x}|)}{\frac{4}{3}\pi r_0^3} 
\nonumber \\
& & - \frac{\int d^3x  
\,~ \overline{S}^{(1)}  \vec{\nabla} |\vec{x}| 
\cdot 
\vec{\nabla} \epsilon_{(1)}
 \, \delta(r_0-|\vec{x}|)}{\frac{4}{3}\pi r_0^3}
\label{3.1}
\end{eqnarray}
where we have neglected the suffix $\Sigma_{A_0}$.
If we now change coordinate system and we move to the spherical
coordinates $(r, \theta, \phi)$ where
\beq
\vec{\nabla} f(r, \theta, \phi)=\frac{\partial f}{\partial r} \hat{e}_r+
\frac{1}{r}
\frac{\partial f}{\partial \theta} \hat{e}_\theta
+\frac{1}{r \sin \theta}\frac{\partial f}{\partial \phi} \hat{e}_\phi
\eeq
we obtain the following result
\begin{eqnarray}
\langle \tilde{S} \rangle_{A_0, r_0}-\langle S \rangle_{A_0, r_0}&=&
\frac{\int \, d\phi \, d\theta \, dr \,   r^2 \sin \theta
\,~ \overline{S}^{(1)}
\, \theta(r_0-r)}{\frac{4}{3}\pi r_0^3} 
\frac{\int \, d\phi \, d\theta \, dr  \, r^2 \sin \theta  
\,~ \frac{\partial}{\partial r}\epsilon_{(1)}
 \, \delta(r_0-r)}{\frac{4}{3}\pi r_0^3} 
\nonumber \\
& & - \frac{\int \,  d\phi \, d\theta \, dr \,  r^2 \sin \theta  
\,~ \overline{S}^{(1)} \frac{\partial}{\partial r} \epsilon_{(1)}
 \, \delta(r_0-r)}{\frac{4}{3}\pi r_0^3}\,.
\label{3.3}
\end{eqnarray}
If $\bar{S}^{(1)}\neq 0$ and we take a radius of integration $r_0$ which is small compared with the typical size of
our cosmological perturbations $r_* \sim 100$ Mpc, the integral of such
first order perturbation is non zero and, as said,   
the gauge dependence will be subleading with respect to this first order contribution.

Let us now consider a region of integration with $r_0>r_*$
and make the hypothesis that we can set to zero the integral of first order 
quantities. Then, following \cite{CAN}, we connect our fluctuations with  
Gaussian primordial ones with zero mean and we use ensemble average   
(\cite{Ruth,LL}) to calculate the expectation value of our averages.
Following the standard 
notation we denote this additional average by an over-bar.

Let us consider the momentum expansion of a general first order
perturbation $\sigma^{(1)}$ in the form
\be
\sigma^{(1)}(\vec{x}) = \frac{1}{(2 \pi)^{3/2}} \int d^3 k \, \e^{i
\vec{k}\cdot \vec{x}} \sigma^{(1)}_k E(\vec{k}) \,.
\label{3.9}
\ee
Using such expansion and the properties mentioned above in 
the Eq.(\ref{3.3}) the first order integrals disappear and one obtains
\begin{eqnarray}
\langle \tilde{S} \rangle_{A_0, r_0}-\langle S \rangle_{A_0, r_0}
&=& -\left(\frac{4}{3}\pi r_0^3\right)^{-1} \frac{1}{(2 \pi)^{3}} \int \,  d\phi \, d\theta \, dr \,  
r^2 \sin \theta  
\,~ \left[\int d^3 k_1 d^3 k_2 e^{i
\vec{k}_1 \cdot \vec{r}} 
\right. \nonumber \\
& & \left.
\overline{S}^{(1)}_{k_1} E(\vec{k}_1) 
\,i k_2  \cos \Omega \, 
e^{i\vec{k}_2 \cdot \vec{r}}
\epsilon_{(1)k_2}\,E(\vec{k}_2)  \right] \delta(r_0-r)
\label{65}
\end{eqnarray}
where $\Omega$ is the angle between $\vec{k}_2$ and
$\vec{r}$.
If our fluctuations are statistically homogeneous, then 
$E$  is a unit random variable satisfying $E^*(\vec{k})=E(-\vec{k})$ and
\be
\overline{E(\vec{k}_1) E(\vec{k}_2)}=\delta(\vec{k}_1+\vec{k}_2).
\label{PropRandomVar}
\ee
It follows that the ensemble average of Eq.(\ref{65}) gives
\begin{eqnarray}
\overline{\left(\langle \tilde{S} \rangle_{A_0}-\langle S \rangle_{A_0}
\right)} &=&  -\left(\frac{4}{3}\pi r_0^3\right)^{-1}
 \frac{1}{(2 \pi)^{3}} \int \,  d\phi \, d\theta \, dr \,  
r^2 \sin \theta  
\,~ \left[\int d^3 k \frac{\overline{S}^{(1)}_{k}}{S^{(0)}} \,
i k \, \right. \nonumber \\
& & \left. 
\cos \tilde{\Omega}\,
\epsilon_{(1)k} \right] \delta(r_0-r)
 \,\,,
\end{eqnarray}
where $\tilde{\Omega}$
is the angular separation between the vector $\vec{k}$ and
$\vec{r}$.
By considering the 
spatial integral inside the momentum integral, 
it is easy to check, by symmetry arguments, that the spatial integral vanishes and
the gauge dependence goes to zero as a consequence of the assumptions made. 
This is not surprising, the assumption that our average
over a limited region may be replaced with an ensemble average 
is reasonable only if we perform the integration on a region which is
much bigger than the scale of our inhomogeneities. In this case the gauge
dependence should, indeed, vanish~\footnote{Actually, if one applies an ensemble average 
directly to Eq.(\ref{3.3}),
using Eqs.(\ref{3.9}) and (\ref{PropRandomVar}) and without setting to zero the integral of first
order quantities, the first term on the r.h.s. can give a non-vanishing 
contribution which is however negligible in the limit $r_0 \gg r_*$.}.

\section {Conclusion}
\label{Sec5}
\setcounter{equation}{0}

This paper is focused on different issues associated with the evaluation 
of the backreaction effects of the classical (quantum) inhomogeneities, as the residual gauge 
dependence of the averaging prescription introduced in \cite{GMV1,GMV2} 
and the choice of the observers 
with respect to whom we evaluate the effects. 

In Sect.\ref{Sec3} we have described how to construct a scalar field $A(x)$,
homogeneous in a given gauge,  
which defines the hypersurface on which we want to perform the average thus introducing  the tools needed to have a correspondence between the standard average prescription \cite{Buchert} and the one introduced in \cite{GMV1,GMV2}. 
Such a scalar will identify the observers with respect to whom the average is performed.
The physical properties of such observers are independent from the zero mode 
of the scalar field and from the possible choice of gauge.
We have a {\it gauge invariant way to select the observers}.
As a key example we gave the scalar and the velocity vector which identify the observers 
in geodesic motion with respect to the perturbed FLRW space-time.

The results obtained in Sect.\ref{Sec3} can 
also be used for the evaluation of the quantum cosmological backreaction.
In fact, the classical averages performed over all 3-dimensional space can be replaced
by quantum expectation values
using the correspondence illustrated in \cite{GMV1}. 

In Sect.\ref{Sec4} we have considered and formalized 
the residual gauge dependence 
associated with the averaging prescription (introduced in \cite{GMV1,GMV2}) 
when the average is performed on a finite volume.
Let us recall again the main results: first, the result (\ref{4.4}) 
is independent
from the particular window function used. 
Second, we confirm (as already stated in \cite{GMV1}) that 
the gauge dependence goes to zero 
for averages over all the 3-dimensional space. 
Third, regardless of  the averaged scalar $S$,  the gauge dependence 
is always at least of the second order.
And, in the end, if $\bar{S}^{(1)}=0$ the average is gauge invariant 
up to second order.

 
\section*{Acknowledgements}

I wish to thank Maurizio Gasperini and Gabriele Veneziano for several discussions
and comments on the manuscript, Adam Christopherson for comments on the 
manuscript and Julien Larena for discussions on the ensemble average. 
I was supported by the GIS ``Physique des Deux Infinis'' during the first stage of this work.


\end{document}